# Complementary Skyrmion Racetrack Memory with Voltage Manipulation

Wang Kang, *Member, IEEE,* Chentian Zheng, Yangqi Huang, Xichao Zhang, Yan Zhou, Weifeng Lv and Weisheng Zhao*, *Senior Member, IEEE*

*Abstract*—Magnetic skyrmion holds promise as information carriers in the next-generation memory and logic devices, owing to the topological stability, small size and extremely low current needed to drive it. One of the most potential applications of skyrmion is to design racetrack memory (RM), named Sk-RM, instead of utilizing domain wall (DW). However, current studies face some key design challenges, e.g., skyrmion manipulation, data representation and synchronization etc. To address these challenges, we propose here a complementary Sk-RM structure with voltage manipulation. Functionality and performance of the proposed design are investigated with micromagnetic simulations.

## I. INTRODUCTION

MAGNETIC skyrmions are particle-like topologically stable spin configurations, which have been observed in both bulk ferromagnets and magnetic thin films [1-3]. A skyrmion is featured by the skyrmion number (Q), which is the topological number in the planar system. In general, we have Q = +1 (or Q = −1) for a skyrmion (or anti-skymion) in a sufficiently large area and it keeps a constant as the skrymion continuously deforms. The sign of Q depends on the spin direction at the core and tail of the spin textures [4, 5]. Since the first observation in 2009 [1], skyrmion has been intensively investigated in both theory and experiment because of the appealing merits, such as topological stability, small size, high mobility and extremely low depinning current [6-8]. It has been widely considered as a promising information carrier for building the next-generation spintronic memory and logic devices. One of the most potential applications of skyrmion is to build racetrack memory (RM) instead of utilizing domain wall (DW) [4, 8]. As skyrmion has smaller size (down to several nanometers) and lower depinning current density (down to $10^6$ A/m$^2$) compared with DW (with size > 30 nm and depinning current density > $10^{10}$ A/m$^2$) [4],
the skyrmion-based RM (Sk-RM) is expected to achieve higher package density and lower power consumption in comparison with the DW-based RM (DW-RM).

Until now, the fundamental functions and characteristics of skyrmion, including skyrmion creation/annihilation, motion and detection etc, have been widely studied [7-9]. However, the application of skyrmion in RM design is still open and lots of design challenges are still remaining, e.g., data representation, manipulation and synchronization. In DW-RM, binary data bits are encoded in a magnetic nanowire by a train of spin-up or spin-down magnetic domains separated by DWs [10]. The train of domains can be moved rhythmically in the nanotrack under a driving current and the binary data bits can be detected by distinguishing the spin orientations of the domains. However, data representation in the Sk-RM is a key problem. A typical approach is to represent the binary data bits with the presence or absence of a skyrmion [8]. In specific, let the presence of a skyrmion represent data bit "1" and the absence of a skyrmion represent data bit "0" or vice versa. Unfortunately, this method is not robust for moving and reading a long sequence of data bits in a nanotrack, especially when taking into consideration the process variations and thermal fluctuations. For example, given a data sequence composed of all "0" bits (absence of skyrmions), it is hard to determine how many bits transmitted in the nanotrack, because the spacing between two skyrmions may be variable in practice. In addition, clock synchronization is a big challenge, as the location of data bit "0" is unpredictable.

In this letter, we propose a complementary Sk-RM structure to address the above-mentioned design challenges. With the aid of voltage manipulation, both bits "0" and "1" are represented in the form of skyrmions and are moved synchronously. In addition, skyrmion detection stability can be greatly improved. Using micromagnetic simulations, we successfully validate the functionality of the proposed complementary Sk-RM structure.

## II. COMPLEMENTARY SK-RM STRUCTURE

Fig. 1 shows the schematic of the proposed complementary Sk-RM structure, which is composed of five parts: a write head for skyrmion creation, two read heads for skyrmion detection, a voltage controlled y-junction for skyrmion selection (i.e., data encoding), two (say L and R) nanotracks for skyrmion motion as well as peripheral CMOS circuits for generating the write current ($I_{write}$), driving current ($I_{drive}$) and detection current ($I_{detect}$). The skyrmion creation, motion and detection have

Manuscript received February 29, 2016. This work was supported in part by the China Postdoctoral Science Foundation (2015M570024), and the National Natural Science Foundation of China (61501013 and 61571023).

Wang Kang an Weifeng Lv are with the Spintronics Interdisciplinary Center (SIC) and School of Computer Science and Engineering, Beihang Univeristy, Beijing, 100191, China (e-mail: wang.kang@buaa.edu.cn).
Xichao Zhang and Yan Zhou are with the Department of Physics, University of Hong Kong, Hong Kong, China (e-mail: yanzhou@hku.hk).
Yangqi Huang, Chentian Zheng and Weisheng Zhao are with Fert Beijing Institute, SIC, Beihang Univeristy, Beijing, 100191, China (corresponding author: Weisheng Zhao; phone: +86-010-82314875; fax: +86-010-82339374; e-mail: weisheng.zhao@buaa.edu.cn).

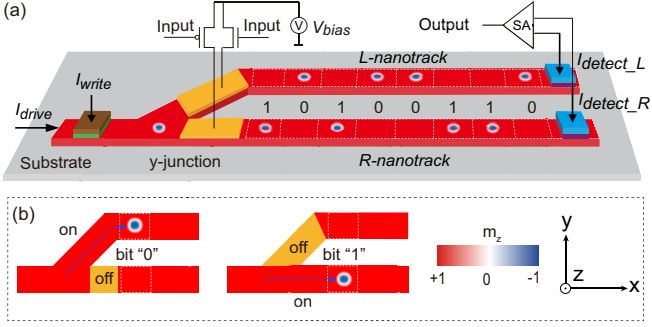

Fig. 1. (a) Schematic of the proposed complementary Sk-RM structure; (b) data representation and encoding under the voltage controlled y-junction.

been widely investigated in literature [7-9], therefore they are omitted in the letter. Here we concentrate mainly on the complementary Sk-RM structure and its implementation. The voltage-controlled y-junction is the key component in our proposed complementary Sk-RM and the other components are similar to the typical Sk-RM structure. The two output (L and R) branches of the y-junction are controlled with a bias voltage by modulating the perpendicular magnetic anisotropy (PMA, the details can refer to [11, 12]). If the (positive) voltage is on, then the corresponding branch of the y-junction will have an energy barrier (due to the PMA difference) to prevent the skyrmion flowing through; otherwise, the skyrmion can move freely through the branch. Here the energy barrier is set to be higher than the driving force of the driving current. With the aid of the voltage controlled y-junction, we can drive the skyrmion into either the L- or R-nanotrack depending on the input data bit (see Fig. 1(b)). For example, if the input data bit is "0", then we switch on the L-branch (adding zero voltage) whereas switch off the R-branch (adding positive voltage) of the y-junction, consequently, the skrymion will move into the L-nanotrack through the L-branch of the y-junction; otherwise, if the input data bit is "1", the skyrmion will move into the R-nanotrack through the R-branch of the y-junction. It should be noted that when a skyrmion is driven into one of the nanotracks, a ferromagnetic state (no skyrmion) will also be driven into the other nanotrack at the same time, as the driving currents of both the nanotracks are simultaneously on. In this case, we can form a complementary data representation and the data bit can be detected by distinguishing the skyrmion location (in either the L-nanotrack or the R-nanotrack). Below we illustrate the full data flow of the proposed complementary Sk-RM structure. The skyrmions are initially created by the write head, and then they move into one of the nanotracks through the y-junction depending on the input data bit value. After that, the skyrmions (data bits) flow along the nanotracks under the driving current. Finally they reach the read head and be readout.

In the proposed complementary Sk-RM structure, data bit is represented by the selection (L or R) of the nanotracks instead of the presence (or absence) of a skyrmion in one nanotrack. This strategy is more robust for data transmission and detection, as we can always detect a skyimon for either data bit "0" or "1". In addition, the skyrmion detection sensitivity and stability are greatly improved with the complementary structure.

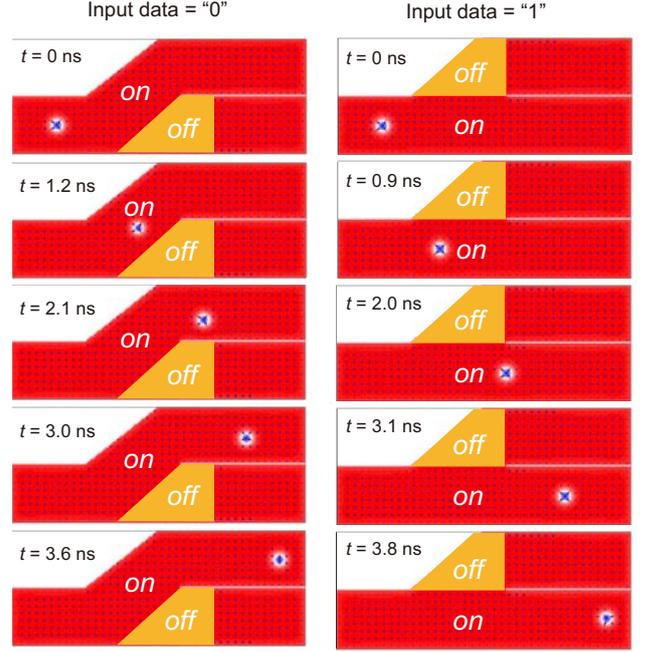

Fig. 2. The micromagnetic simulation results: the top-view of the snapshots of magnetization configurations at various selected times.

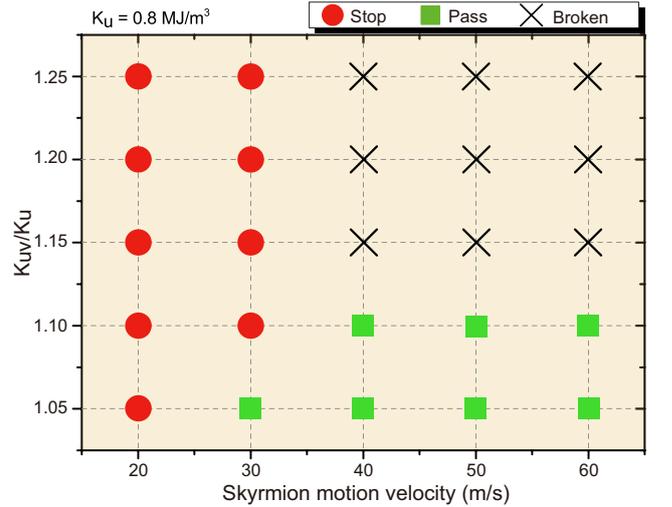

Fig. 3. The working window of the voltage-controlled y-junction at various energy barriers ($K_{uv}$) and driving current densities.

## III. SIMULATION AND DISCUSSION

To validate the functionality as well as to investigate the behavior characteristic of the proposed complementary Sk-RM structure, we performed micromagnetic studies by solving the well-known Landau-Lifshitz-Gilbert (LLG) equation with the spin transfer torques [7, 13], which governs the dynamics of the magnetization $\boldsymbol{m}$ at each lattice,

$$\frac{d\boldsymbol{m}}{dt} = -|\gamma|\boldsymbol{m}\times\boldsymbol{H}_{eff} + \alpha\boldsymbol{m}\times\frac{d\boldsymbol{m}}{dt} + u\boldsymbol{m}\times\left(\frac{\partial\boldsymbol{m}}{\partial x}\times\boldsymbol{m}\right) + \beta u\left(\boldsymbol{m}\times\frac{\partial\boldsymbol{m}}{\partial x}\right) \quad (1)$$

where $\gamma$ is the gyromagnetic ratio, $\alpha$ is the Gilbert damping, $u = \gamma\hbar jP/2eM_s$, $\hbar$ is the reduced Planck constant, $j$ is the driving current density flowing in the $-x$ direction, $P$ is the spin polarization, $e$ is the electron charge, $M_s$ is the saturation magnetization, and $\beta$ is the non-adiabatic factor [14], $\boldsymbol{H}_{eff}$ is the effective field including the contributions of Heisenberg exchange, Dzyaloshinskii-Moriya interaction (DMI), magnetic anisotropy as well as demagnetization field. In the simulations, the nanotrack is a 0.4-nm-thick Cobalt on a Platinum substrate. The width and length of the y-junction are 80 nm and 200 nm. The spacing between the two nanotracks is 2 nm (it prefers to be as narrow as possible to achieve high package density). We adopted the following material parameters (default) in our simulations [7, 13]: exchange stiffness $A = 15$ pJm$^{-1}$, Gilbert damping $\alpha = 0.3$, saturation magnetization $M_s = 580$ kAm$^{-1}$, spin polarization $P = 0.4$, DMI constant $D = 3$ mJm$^{-3}$, PMA of the nanotracks are $K_u = 0.8$ MJm$^{-3}$, PMA of the y-junction with positive voltage is $K_{uv} = 0.88$ MJm$^{-3}$. The relationship between the skyrmion motion velocity and driving current can be obtained by solving the Thiele equation [8, 15],

$$\boldsymbol{F} + \boldsymbol{G} \times (\boldsymbol{v} - \boldsymbol{u}) + \boldsymbol{D}(\alpha\boldsymbol{v} - \beta\boldsymbol{u}) = 0 \qquad (2)$$

where $\boldsymbol{F}$ is the pinning forces applied on skyrmions, $\boldsymbol{G}$ is the gyromagnetic coupling vector, $\boldsymbol{D}$ is the dissipation matrix, $\boldsymbol{v}$ and $\boldsymbol{u}$ are the motion velocities of the skymions and conduction electrons respectively.

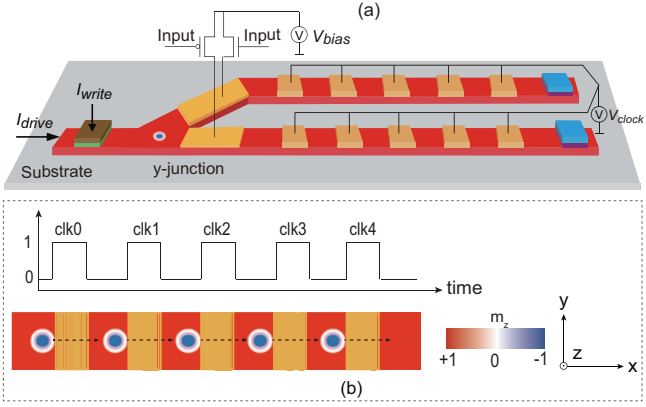

Fig. 4. Illustration of the clock synchronization with voltage manipulation. (a) The schematic, and (b) skyrmion motion under the clock signal.

Fig. 2 shows the top-view of the magnetization evolutions of the y-junction. As can be seen, the skyrmion can be selectively driven into the L- or R-nanotrack using the voltage-controlled y-junction. Here the driving current density is $10^{12}$ A/m$^2$ and the skyrmion motion velocity is ~50 m/s. It is worth noting that there is a design tradeoff between the skyrmion motion velocity (or driving current density) and the energy barrier ($K_{uv}$) of the voltage-controlled y-junction. More specifically, if the driving force of the current is higher than the energy barrier of the y-junction, the skyrmion selection (i.e., data encoding) may fail, as the skyrmion will move across the energy barrier of the y-junction and move into the wrong nanotrack, even though the branch of the y-junction is expected to be off. To make matters worse, the skyrmion may be broken due to the non-equilibrium competition between the driving current and the effective magnetic anisotropy [4, 7]. Fig. 3 shows the working window of the y-junction at various energy barriers ($K_{uv}$) and skyrmion motion velocities. The mechanism of voltage-controlled PMA effect can also be utilized to improve the clock synchronization of the Sk-RM. As shown in Fig. 4(a), we can add a series of voltage-controlled gates along the nanotrack and the on/off voltages of the gates are controlled by a clock signal. When the voltages of the gates switch on, the skyrmions stop and are pinned at the region of the gates; otherwise, when the voltages of the gates turn off, the skyrmions are depinned and continue to move until the next gate. In this configuration, the skyrmions in the nanotracks can move step-by-step under the control of the clock signal (see Fig. 4(b)).

## IV. CONCLUSION

In this letter, we proposed a complementary Sk-RM structure. With the aid of a voltage-controlled y-junction, the skyrmions can be selectively driven into two different nanotracks, forming complementary data representation. In the proposed Sk-RM, both data bits "0" or "1" are represented with the presence of a skyrmion, therefore improving the data robustness and clock synchronization. Functionality of the proposed complementary Sk-RM structure is validated with micromagnetic simulations. This work gives guidelines for the design and development of advanced skyrmion-based racetrack memory.